\documentclass{sig-alternate}
\usepackage{subfigure}
\usepackage{mathrsfs}
\usepackage{graphicx}
\usepackage{enumitem}
\newtheorem{definition}{Definition}

\begin{document}
%
\conferenceinfo{WOODSTOCK}{'97 El Paso, Texas USA}

\title{Predicting Trends in Social Networks via Dynamic Activeness Model}

%
%
%
%
%

\numberofauthors{1} 
%
\author{
%
%
\alignauthor
\begin{tabular}{cccc}
Shuyang Lin$^\dag$ &
Xiangnan Kong$^\dag$ &
Philip S. Yu$^{\dag*}$\\
\end{tabular}\\
\begin{tabular}{ccc}
      \affaddr{$^\dag$Department of Computer Science} &  &\affaddr{$^*$Computer Science Department}\\
      \affaddr{ University of Illinois at Chicago} & &  \affaddr{King Abdulaziz University}\\
			\affaddr{ Illinois, USA} & &  \affaddr{Jeddah, Saudi Arabia}
\end{tabular}\\
\email{\{slin38,xkong4,psyu\}@uic.edu}
}


\maketitle
\begin{abstract}
With the effect of word-of-the-mouth, trends in social networks are now playing a significant role in shaping people's lives. 
Predicting dynamic trends is an important problem with many useful applications. 
There are three dynamic characteristics of a trend that should be captured by a trend model: intensity, coverage and duration.
However, existing approaches on the information diffusion are not capable of capturing these three characteristics. 
In this paper, we study the problem of predicting dynamic trends in social networks.
We first define related concepts to quantify the dynamic characteristics of trends in social networks, 
and formalize the problem of trend prediction.
We then propose a Dynamic Activeness (DA) model based on the novel concept of activeness,
and design a trend prediction algorithm using the DA model.
Due to the use of stacking principle, 
we are able to make the prediction algorithm very efficient.
We examine the prediction algorithm on a number of real social network datasets, 
and show that it is more accurate than state-of-the-art approaches.
\end{abstract}


\section {Introduction}
Online social networks have become increasingly important for interpersonal communication and information sharing. 
Trends in online social networks now have large impacts on people's lives.
Trends are represented by sequences of actions that are taken by users in a social network. 
According to the type of the social network, an action can be posting a blog or sharing a webpage about a certain topic, or joining an online activity.

Predicting the dynamic behavior of trends is an interesting problem with wide applications.
Some examples of such applications are as follows:
\begin{enumerate}
\item  Online video providers may want to predict how many times a video will be played by users in the next month,
so that they can decide the bandwidth needed for the server.
\item  Disease control facilities may want to predict how many people will suffer from a contagion in the following week, so that they can be prepared for an outbreak. 
\item  Manufacturers may want to predict how long an existing product will continue to be popular, 
so that they can decide the most suitable time for the debut of a new model.
\end{enumerate}

The three applications above require the prediction of trends from three different perspectives.
The first example considers the \textbf{intensity} of a trend, which is the volume of actions during a fixed length of time.
The second one focuses on the \textbf{coverage} of a trend, which is the number of people taking the given action during a fixed length of time.
The third one considers the \textbf{duration} of a trend, which is the time span that the intensity or coverage is above a given threshold.

To better explain these three perspectives (intensity, coverage and duration), we show in Figure \ref{fig:example} a toy example of a trend on a social network which contains three users.
Table (b) shows the intensity, coverage and duration that aggregated from actions listed in Table (a).
For example, at 2008, $v_1$ and $v_2$ take 3 and 2 actions,  respectively,  while $v_3$ taking no action, so the coverage (i.e. the number of people taking actions) is 2, and the intensity  (i.e., the total number of actions taken) is 5.
Though the coverage and intensity are correlated with each other, they are not interchangeable in the sense that the corresponding time series are neither similar nor synchronized.
In this example, the maximum value of coverage is reached at year 2009, while the maximum value of intensity is reached at year 2008.
Duration reflects how long the trend lasts.
If we set the threshold to $0$, duration of the trend will be 4 years, from 2007 to 2010. 

\begin{figure}[t]
\centering
\subfigure[\small Number of actions of the trend]{
\begin{tabular}{ l  l  l  l  l  l  }
	\hline
	 & 2007 & 2008 & 2009 & 2010 & 2011\\\hline 
	$v_1$&	1&3&2&1&0\\
	$v_2$&	0&2&1&0&0\\
	$v_3$&	0&0&1&1&0\\ \hline
\end{tabular}}\\
\subfigure[\small Coverage, intensity, and duration of the trend]{
\begin{tabular}{ l  l  l  l  l  l } 
	\hline
	 & 2007 & 2008 & 2009 & 2010 & 2011 \\\hline 
	Coverage&  1 & 2 & 3 & 2 & 0 \\ 
	Intensity& 1 & 5 & 4 & 2 & 0 \\ 
	Duration& \multicolumn{5}{c}{2007 - 2010}\\\hline
\end{tabular}}
\caption{An example of trend in a social network}	
  \label{fig:example}
\end{figure}

Based on our observation, 
each of the three perspectives is useful for many real applications.
A trend model should be able to characterize trends from all of these three perspectives.

Though the actions of social network users have been studied in the context of information diffusion models (e.g. the independent cascade (IC) model) \cite{Kempe2003,Goyal2010,Leskovec,Leskovec2007,Chen2009,Chen2010,Gruhl2004,Kossinets2008},
the existing information diffusion models are not suitable for modeling dynamic trends for three main reasons:
\textbf{First}, most of these models assume that the diffusion processes take place in discretized time
and the propagation of information between two nodes always takes one unit of time, 
which does not reflect the real dynamic of trends as time unfolds.
Therefore, they cannot reflect the dynamic nature of intensity and coverage, or the duration of trends.
\textbf{Second}, information diffusion models focus on the visible path of propagation, and they usually assume that the propagation can only occur between a pair of nodes that are directly linked to each other,
while the trend model should focus on predicting the aggregate characteristics of trends, and the path of propagation is not important for the prediction.
Besides, because of the existence of homophily \cite{Aral2009,Shalizi2011}, the propagation through direct links may not be good enough to explain trends in social networks. 
The model of trends should be more flexible with regard to the propagation mechanism.
\textbf{Third}, information diffusion models focus on the prediction on the individual user level, but not on the trend level.
As a result, they allow the probability of influence to be different between each pair of users, 
but assume that the probability remains the same for all the trends.
This makes them not suitable for predicting trends based on different properties of trends.

In this paper, we formally define the three dynamic characteristics of a trend (intensity, coverage and duration), and the problem of trend prediction.
We introduce a novel concept of activeness, which reflects a user's interest toward the given trend at a given point of time.
The dynamic nature of activeness enables us to model the dynamic characteristics of trends.
Due to the introduction of activeness, a more flexible propagation mechanism is made possible, so that the correlation as well as influence through direct links can be captured in our model. 
We propose a Dynamic Activeness (DA) model based on the concept of activeness.
Each component of the model is built on observations on real trends,
and the model is capable of capturing all the three dynamic characteristics of trends. 
We design a trend prediction algorithm based on the DA model.
For each trend, the parameters of the model are learned specifically from the history data of that trend.
The learned model can then be used to predict the dynamic characteristics of the trend in the future.
For the efficiency consideration, 
we identify the stacking principle and utilize it to transform the effect of the sequence of actions to the sum of the individual effects by each single action in isolation.
This makes the prediction based on the DA model have a similar computational complexity to the IC model. 
We show the performance of the DA model on real trends in social networks.

\section{Preliminaries}
\subsection{Notations and Definitions}
Let $G = (V,E)$ be a social network, in which $V = \{v_1, \cdots,$ $v_n\}$ is the set of nodes and $E\subseteq V \times V$ is the set of edges. 
We consider the network to be static, since the evolution of networks is much slower than that of the trends.
During the time span of a given trend, the change of the network is negligible.

A trend on the social network $G$ is defined as follows:
\begin{definition} {\bf Trend} 
	A trend $S = [(v_1,t_1),\ldots,(v_m,t_m)]$ on the social network $G$ is a chronological sequence that consists of a given type of actions in $G$,
	where $t_i \le t_j (\forall 0\le i<j \le m$) and $v_i \in V (\forall i\in\{1, \cdots m\})$. 
	An element $(v_i, t_i)$ in $S$ corresponds to an action of that type taken by the node $v_i$ at time $t_i$. 
\end{definition}

For simplicity of notation, we denote the time sequence of actions in trend $S$ as $T(S) = [t_1,\ldots t_m]$, 
and denote the subsequence of $S$ that consists of all the actions taken by node $v$ as $S^v = [(v,t_{v,1}),\ldots,(v,t_{v,m_v})]$, where $(v,t_{v,i})\in S (\forall 1\leq i\leq m_v)$.
We also use $S_{t}$ to denote the prefix of $S$ which contains all the actions taken before the time point $t$, i.e., $S_{t} = [(v_i,t_i): (v_i,t_i) \in S, t_i\leq t]$.

Based on the above definition of a trend, we define intensity, coverage, and duration of a trend as follows:

\begin{definition}{\bf Intensity}
The Intensity of a trend $S$ on a time interval $I=[t_{min},t_{max})$ is the number of actions in $S$ that are taken during $I$. 
	Formally, $Intensity(S,I)=|\{(v_i,t_i): t_i \in I \wedge (v_i,t_i) \in S\}|$.
\end{definition}

\begin{definition}{\bf Coverage}
The Coverage of a trend $S$ on a time interval $I=[t_{min},t_{max})$ is the number of nodes in $V$ that take at least one action during $I$.
	Formally, $Coverage(S,$ $I)=|\{v_i: (v_i,t_i)\in S \wedge t_i\in I\}|$.
\end{definition}

\begin{definition}{\bf Duration}
	Let ${\mathcal{I}}= \{I_1,\ldots I_s\}$ be a set of intervals, where $I_i=[t^i_{min},t^i_{max})$. 
	Given a threshold $\theta$, the duration of $S$ on $\mathcal{I}$ is the number of consecutive intervals in  $\mathcal{I}$ that the intensity (coverage) is above $\theta$. 
	Formally, $Duration_{cov}$$(S, \mathcal{I},\theta)=max (j-i+1), 1\leq i\leq j\leq s,\ s.t.\ \forall k, i\leq k \leq j, \ Coverage(S,I_k)>\theta$ and $Duration_{int}$ $(S, \mathcal{I},\theta)=max (j-i+1), 1\leq i\leq j\leq s,\ s.t.\ \forall k, i\leq k \leq j,\ Intensity(S,I_k)>\theta.$
\end{definition}

The intensity quantifies the overall activeness of a trend within a social network.
	The coverage quantifies how broad a trend has impacts in a social network, i.e., the number of nodes involved within a time interval.
	The larger the coverage value of a trend is, the more nodes of the network are affected/involved in the trend.
The duration quantifies how long a trend lasts within the social network.

For the duration, we usually want $I_1,\ldots,I_s$ to be consecutively connected intervals with equal length, i.e., $t^i_{max} = t^{i+1}_{min}\ (\forall i \in \{1,\ldots,s-1\})$ and $t^i_{max} - t^i_{min} = t^j_{max} - t^j_{min}\ (\forall i,j \in\{1,\ldots,s\})$.
We like to point out that, given the intensity and coverage, the duration of trend can be defined in many different ways.
We define it as the largest number of consecutive intervals above the threshold because this definition is most straightforward.
By carefully setting the threshold $\theta$, the definition will accord with the intuitive understanding of the word ``duration''.

The prediction problem of trends is defined as follows:
\begin{definition}{\bf Trend Prediction Problem}
	Given $S_{t_*}$, the prefix of sequence $S$ before time $t_*$, the problem of trend prediction is to predict the intensity, coverage and duration of trend S after time $t_*$.
\end{definition}

Typically, we solve the prediction problem on a set of consecutively connected equal-length intervals ${\mathcal{I}}= \{I_1,\ldots I_s\}$, where $I_i=[t^i_{min},t^i_{max})$ and $t^1_{min} = t_*$. 
The problem is to predict $Coverage(S,I_i)$ and $Intensity(S,I_i)$ for each $I_i\in \mathcal{I}$, and $Duration_{cov}(S, \mathcal{I}, \theta_S)$ or $Duration_{int}(S, \mathcal{I}, \theta_S)$ for a given $\theta_S$.

\subsection{Datasets}
\label{sec:DataSets}
We take our observation and evaluation on two social networks: DBLP co-author network and Twitter user network.

{\bf DBLP co-author network:}  
In this network, the nodes correspond to the authors and the edges correspond to the co-authorship.
The dataset contained 934,672 nodes and 8,850,502 edges.
The trend data are extracted from the DBLP data by detecting terms in the titles of publications. 
In each trend, an action $(v_i, t_i)$ corresponds to a publication of the author $v_i$ at time $t_i$ that contains the given term in the title.
For publications with multiple authors, there is an action for each of the authors.

{\bf Twitter user network:} 
In the network, the nodes correspond to the users and the edges correspond to the who-follows-whom relationships.
The edges are directed from the users that are being followed to the followers.
We randomly crawl a sub-network of Twitter network, which contains 40,906 nodes and 7,829,834 edges.
The trends are defined by hashtags in tweets.
An action $(v_i, t_i)$ in a trend corresponds to a tweet of user  $v_i$ at time $t_i$ that contains the given hashtag.

\section{Dynamic Activeness (DA) Model}
\label{sec:model}

\subsection{Concept of Activeness}
\label{sec:concept_activeness}
The DA model for trend in social network is based on the novel concept of \textit{activeness}. 
For each trend, each node in the social network has an activeness function associated with it. 
The two main aspects of the concept are:
\begin{itemize}[noitemsep,nolistsep,leftmargin=5pt,itemindent=7pt]
\item Activeness of a node is defined as its interest toward the given trend. It is a function of time.
	As time goes by, activeness may increase as a result of information diffusion or social influence,
	or decrease as the node loses interest to the given trend.
\item Activeness decides the frequency of actions taken by the node.
      The higher the activeness of a node is, the more actions it is likely to make in a unit time.
			In this sense, we can also define activeness as the ``action rate'' of a user.
\end{itemize}

Since activeness is dynamic in nature, we are able to design the DA model based on it, so that the model can capture the three dynamic characteristics of trends.
Besides, by using activeness in the model, we are also able to design a more flexible information propagation mechanism, as we will show in Section \ref{sec:propagation}.

\subsection{Framework of DA Model}
\label{sec:frame}
\begin{figure}[thb]
\centering
\includegraphics[width=0.45\textwidth] {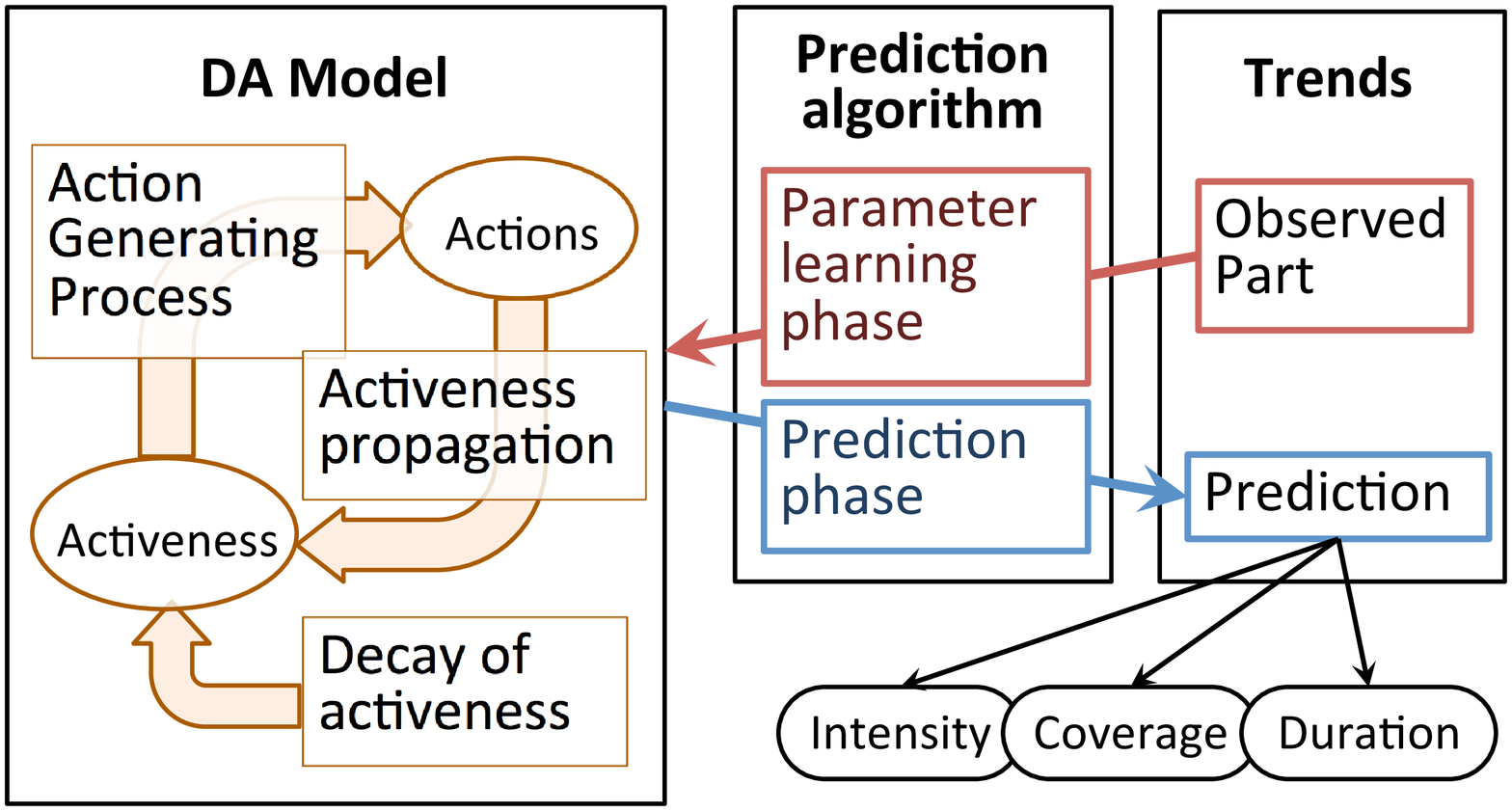}
\caption{Block Diagram of the DA Model}
\label{fig:block}
\end{figure}

As shown in Figure \ref{fig:block}, 
the DA model contains three elements: activeness propagation, decay of activeness and action generating process.
Each of them is based on observations on real trends.
Actions and activeness are connected to each other in the DA model.
On the one hand, actions trigger activeness propagations in the social network. 
Activeness propagation, together with the decay of activeness, decides the activeness of each user at each point of time.
On the other hand, actions are generated by the action generating process which takes the activeness as input.

As shown in Figure 2, the prediction algorithm contains two phases.
In the learning phase, parameters of the DA model are learned from the observed part of trends.
In the prediction phase, the DA model predicts the actions in the future. 
Intensity, coverage and duration of trends can be predicted by aggregating the predicted future actions.


\subsection{Activeness Modeling}
\label{sec:activeness}
For each trend $S$, let $r_{v}(t)$ be the activeness or action rate of node $v$ at time $t$. 
In this section, we discuss the modeling of $r_{v}(t)$.
As shown in the left most box in Figure \ref{fig:block}, the model of activeness contains two parts: activeness propagation and activeness decay.

\subsubsection{Activeness Propagation}
\label{sec:propagation}
Intuitively, as a result of social influence or homophily \cite{Shalizi2011,Budak2011},
the activeness of nodes are correlated with each other.
The closer two nodes are, the larger the correlation is.
Thus, when a node takes an action in a trend, 
we can expect that nodes in proximity to it have larger activeness for the trend than other nodes in the social network.

\begin{figure}[t!]
\centering
\subfigure[]{
\includegraphics[width=0.22\textwidth] {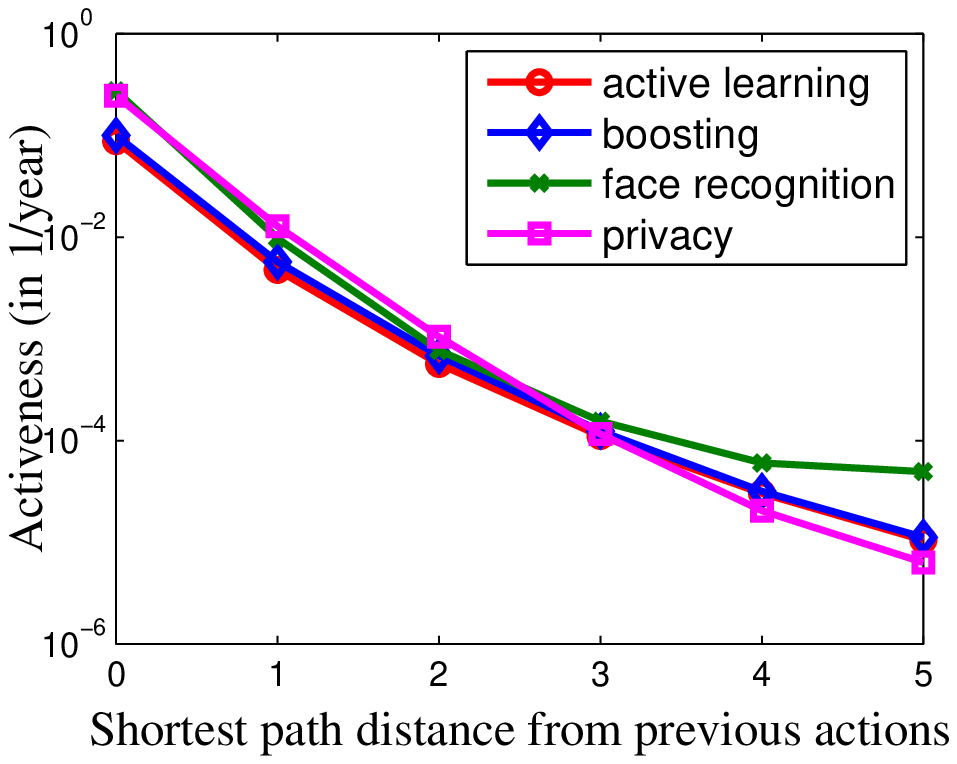}
\label{fig:exct_dist}
}
\subfigure[]{
\includegraphics[width=0.22\textwidth] {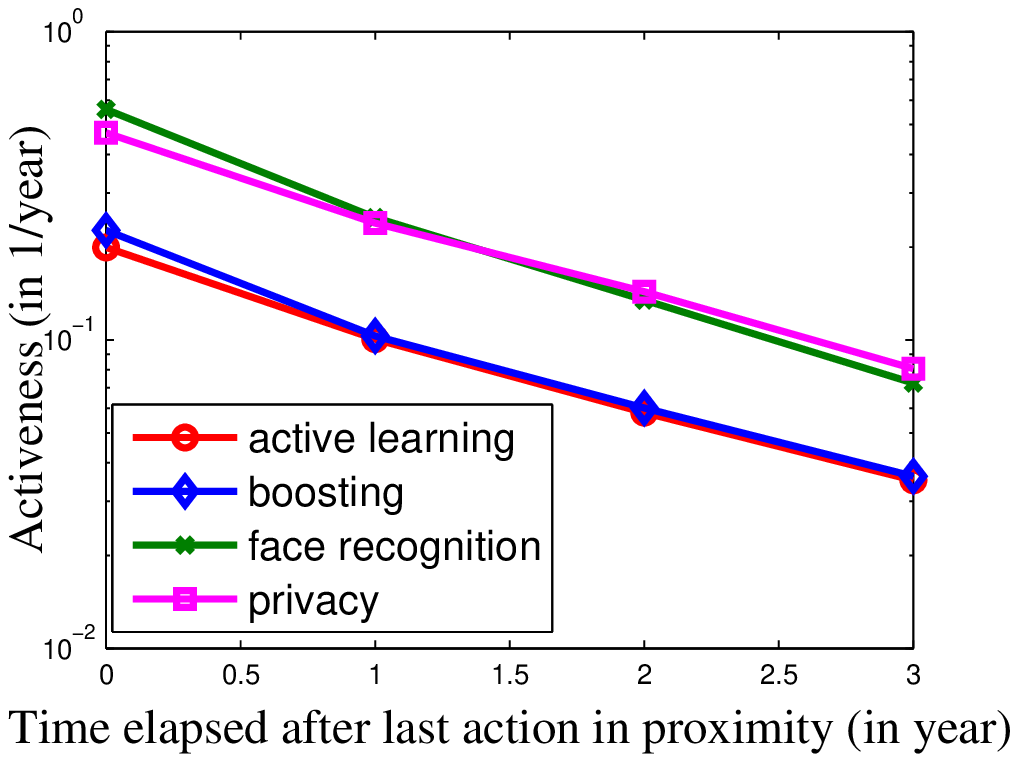} 
\label{fig:exct_delta}
}
\caption{(a)Activeness by different distances to nodes with previous actions. 
	(b)Activeness by the time elapsed since last action in proximity}
\end{figure}

Our observation on the real trends supports this intuition.
Figure \ref{fig:exct_dist} shows the curves of activeness for four trends in DBLP network, i.e., trends about ``boosting'', ``privacy'' etc.
Other trends have similar curves.
For each trend, we plot the average activeness (i.e. the average number of actions per unit time) for nodes with different shortest path distances to nodes with previous actions.
As we can see from the figure, as the distance increases, the activeness of nodes exponentially decreases. 
An important remark is that the information diffusion along direct links is not enough for explaining trends.
Because if we explain trends in that way, all the nodes except for those that are directly linked to the nodes with previous actions should have the same activeness. 
It is also interesting to point out the exponential decreases also fits for action rates of nodes with 0-hop distance (i.e. nodes themselves have previous actions).

Based on this observation, we use the previous actions of a trend to model the activeness of nodes.
Let $prox(u,v)$ be the proximity measurement from $u$ to $v$. 
When an action is taken by node $u$, the increase of activeness of $v$ is proportional to $prox(u,v)$, 
i.e., if $u$ takes an action at time $t_a$, for every node $v$ in the network we have:
\begin{equation}
	\lim_{t\to t_a+} r_v(t) = \lim_{t\to t_a-} r_v(t) + \alpha \cdot prox(u,v)
\label{eq:update}
\end{equation}
where $\lim_{t\to t_a+}r_v(t)$ and  $\lim_{t\to t_a-}r_v(t)$ are the activeness of node $v$ after and before the jump at time $t_a$, and $\alpha$ is the propagation ratio that depends on the trend.

We study two different measurements for proximity. 
The first one uses the shortest path distance from the source node to destination node. 
As we observed in the real trend data, proximity is defined as an exponential decreasing function of shortest distance, i.e., $prox(u,v) = exp(-b\cdot dist(u,v))$, where $b \in \mathcal{R}^+$ and $dist(u,v)$ is the length of shortest path from $u$ to $v$. 
The second proximity measure is based on random walk, which is described as rooted PageRank in \cite{Liben-Nowell2007}. 
To measure the proximity of nodes from a given node $u$, the random walk is started at node $u$. 
At each step, it has a probability of $p$ to return to node $u$, and $1 - p$ probability to move to neighbor nodes. 
The proximity of node $v$ from node $u$ is defined as the stationary probability for $v$.
In both of the measurements, if $v$ is not reachable from $u$, we have $prox(u,v)=0$.

While information diffusion models define the influence between nodes along the edges,
the ``propagation'' of activeness captures a more general sense of correlation between activeness of nodes, 
rather than the process of information diffusion. 
Besides, it is different from information diffusion model in how it is parameterized.
The parameter $\alpha$ of the activeness propagation depends only on the trend, but not on the node that takes the action.
To the contrary, the information diffusion models usually have diffusion probabilities with individual edges as parameters, and the diffusion probabilities are constant for all the trends.
We parameterize the activeness propagation in a different way for two reasons:
First, it is simply not practical to make the parameters depend on trends and edges at the same time, 
since there will be not enough actions to be used for the learning of each parameter.
Second, the main purpose of the proposed trend model is to predict the aggregate characteristics of different trends,
so it is more meaningful to bind the parameter with the trends instead of the edges.

\subsubsection{Decay of activeness}
\label{sec:decay}
Intuitively, if a user is not exposed to any new information or influence from a certain trend, nor does he create any new content that belongs to that trend,
the user's interest to that trend will gradually decay.
In other words, the activeness of a node spontaneously decays if there is no new action taken by nodes in proximity to it.
This spontaneous decay is observed from real trends.

Figure \ref{fig:exct_delta} shows the average activeness for nodes as time progresses since last action in proximity. 
Let $R(v,k) = \{u\in V| sp(u,v) = k\}$ represent the set of nodes that are $k$ hops away from $v$,
where $sp(u,v)$ is the shortest path distance from $u$ to $v$.
The X-axis of the figure is the time elapsed since last action taken by nodes in $R(v,1)$. 
The Y-axis of the figure is the average activeness.
As shown in the figure, activeness decreases when the node is not exposed to a new action. 
Decrease of activeness can roughly be regarded as exponential. 
(We only show the case of $k=1$ and the curves for four trends in this figure. 
But actually we have done the same test for different $k$ values and for different trends, 
and the other curves are similar.)

According to the observation, we introduce an exponential decrease to the activeness model. 
For each interval $[t_0,t_1)$ when $r_v(t)$ is not increased by the activeness propagation, $r_v(t) = r_v(t_0)e^{-(t-t_0)/\tau}$ for any $t\in [t_0,t_1)$. $\tau$ is a rate of the activeness decay, 
For similar reasons as to the parameter $\alpha$, $\tau$ depends on the trend, but not depends on the node or the edge.


\subsubsection{Summary of Activeness Model}
\label{sec:combine}
Combining the two parts discussed in Section \ref{sec:propagation} and Section \ref{sec:decay},
$r_v(t)$, the activeness of node $v$ at time $t$, is given by: 
%
\begin{equation}
r_v(t) = \alpha \sum_{(v_i,t_i)\in S_t}(prox(v_i,v)\,e^{-(t-t_i)/\tau}) + r_v(t_0)e^{-(t-t_0)/\tau}
\label{eq:r}
\end{equation}
where $t_0$ is the start time of the trend. 
We set $r_v(t_0)$  to a small value $\epsilon$. 
$r_v(t)$ is discontinuous at time points when there is a new action taken by nodes in $R(v)$.
In each interval between those discontinuous points, $r_v(t)$ is subject to an exponential decay. 

\subsection{Action Generating Process}
\label{sec:poisson}
In this section, we discuss the generating process for actions.
As we mentioned in Section \ref{sec:concept_activeness}, the activeness of a node serves as the action rate in the generating process. 
With the assumption that time points of actions are conditionally independent, given the activeness function,
the generating process for the sequence of time points is a non-homogeneous Poisson process. 
This assumption keeps the model simple. 
We will first show that this assumption is reasonable for the action generating process,
and then discuss the detail of the generating process under this assumption.

The verification is based on the fact that inter-action time of a homogeneous Poisson process, i.e. a Poisson process with constant action rate, follows exponential distribution.
The sequence for each node usually contains only a few actions in a trend, which are not sufficient for the analysis. 
Instead, we use the global time sequence of all the actions in a trend. 
As a property of Poisson processes, the sum of two Poisson processes is also a Poisson process.
For our case, if each individual action sequence is generated by a Poisson process, the global sequence is also generated by a Poisson process. 
The action rate function of this sequence is the sum of all nodes' action functions, i.e., $r(t) = \sum_{v\in V}r_v(t)$. 
Since $r(t)$ reflects the global activeness of all nodes, it will not change too much in a short term. 
Therefore, the distribution of the inter-action time should roughly be an exponential distribution. 

\begin{figure}[t]
\centering
\subfigure[\small Apple]{\includegraphics[width=0.15\textwidth] {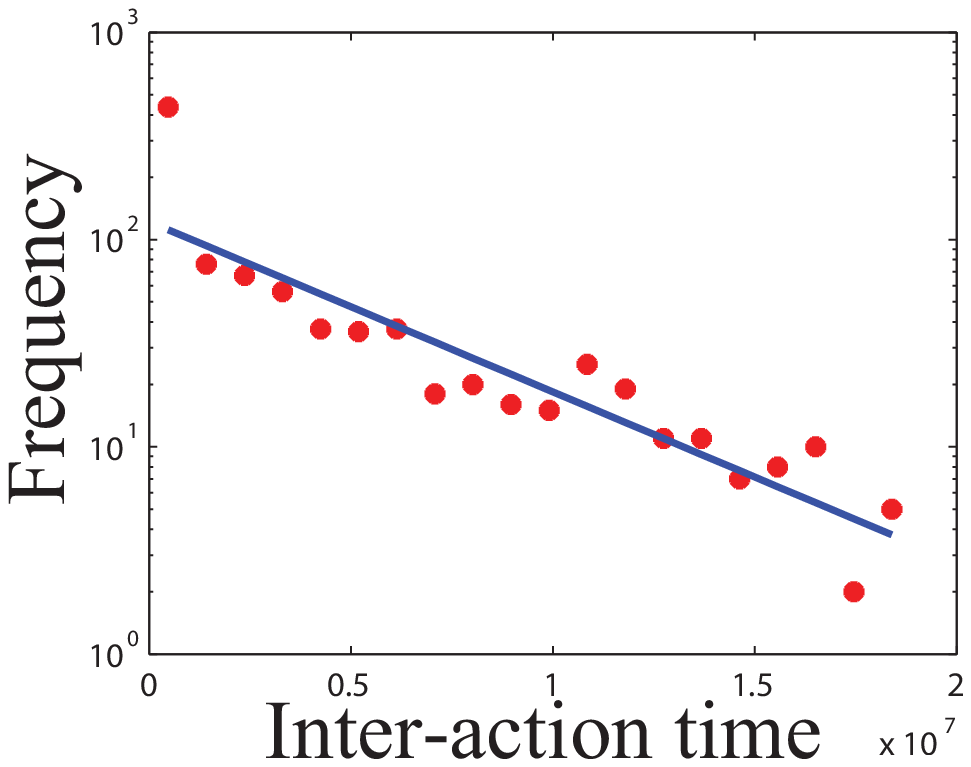}}
\subfigure[\small Android]{\includegraphics[width=0.15\textwidth]{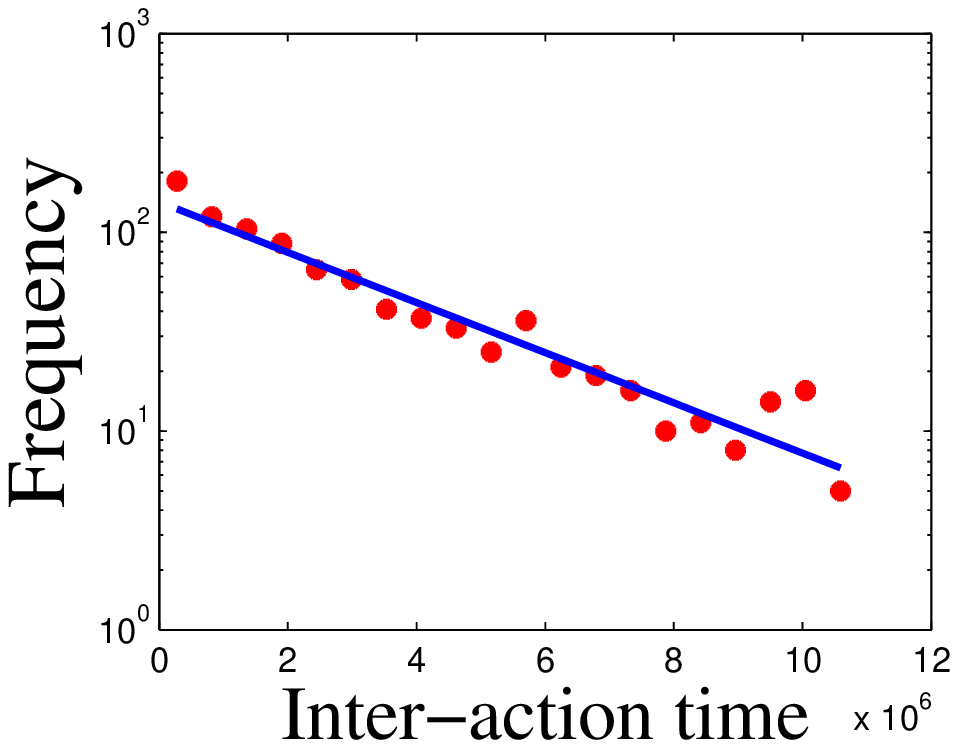}}
\subfigure[\small HTML5]{\includegraphics[width=0.15\textwidth] {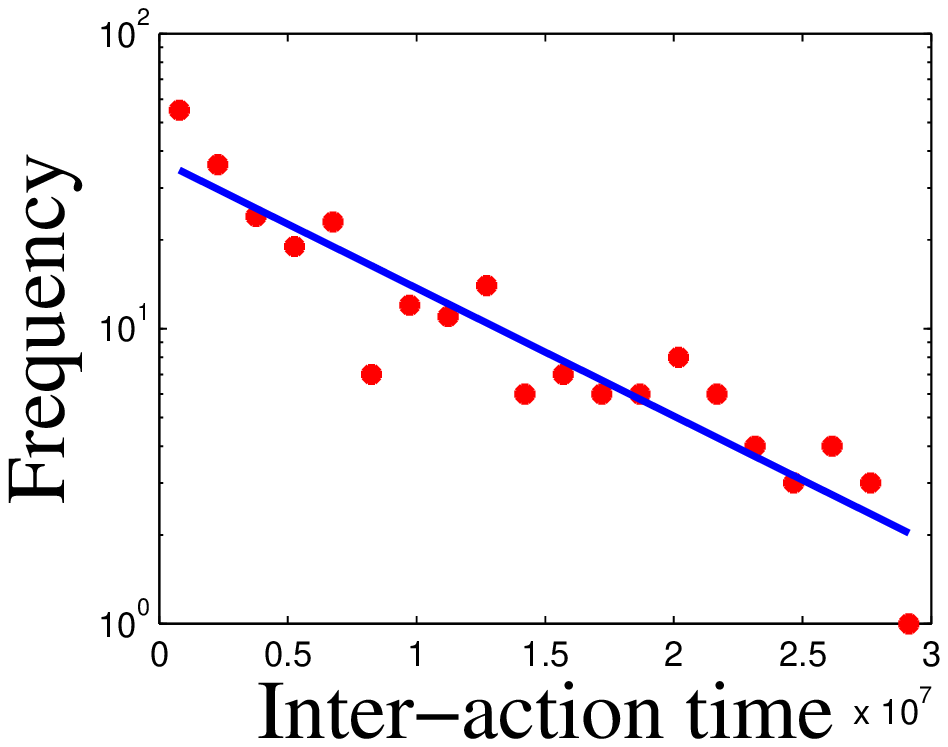}}
\caption{Distribution of inter-action time for three real trends. Other trends have similar curves.}
\label{fig:exp_test}
\end{figure}

We plot the inter-action time distribution for three trends in the Twitter network in Figure \ref{fig:exp_test}.
For each of the trends, we show the distribution of inter-action time during a week period when the trend is popular in the Twitter network.
We divide the inter-action time into 20 bins according to the length, and plot the frequency of each bin.
As showed by these figures, the inter-action time fits exponential distributions quite well.  
Thus, we conclude that the independent assumption is reasonable.

Under the conditional independent assumption, we derive the process of generating actions as follows.
For a non-homogeneous Poisson process, 
the number of actions taken by this node in any time interval $[t',t)$ follows a Poisson distribution:

\[P[(|S^v_{t}|-|S^v_{t'}|) = k] = \frac{e^{-\int_{t'}^{t} r_v(t)\,dt}(\int_{t'}^{t} r_v(t)\,dt)^k}{k!}\]
where $|S^v_{t}|$ is the number of actions taken by node $v$ before time point $t$.

Suppose we are now at the time point $t'$ and want to generate the next time point after $t'$.
By taking derivative with respect to t, we can get the probability density function for the waiting time until next action:

\begin{align}
f_{v,t'}(t) 
&= r_v(t)\cdot exp( \int_{t'}^{t} r_v(t)dt)
\label{eq:waiting_time}
\end{align}

We then can generate the next time point in the sequence by drawing from the distribution of the waiting time.



%
%

\section {Prediction Algorithm}
\label{sec:prediction}
In this section, we show prediction algorithm based on the proposed DA model.
The algorithm contains two phases: the parameter learning phase and the prediction phase.
In the first phase, parameters for each trend are learned from the part of the trend before $t_*$ by maximum likelihood estimation. 
In the second phase, we use the learned model to predict the future trend sequence after $t_*$.

\subsection{Parameter Learning Phase}
For each trend, two parameters in the DA model need to be learned: the proportionality factor $\alpha$ in activeness propagation and mean lifetime $\tau$ of activeness decay. 
We use maximum likelihood estimation for the parameter learning.

The likelihood function is given by:
\begin{equation}
L(\alpha,\tau) = \prod_{v\in V} f(T(S^v_{t_*}), |S^v_{t_*}|;\alpha,\tau)
\label{eq:L1}
\end{equation}
where $T(S^v_{t_*})$ is the time sequence for node $v$'s actions before time $t_*$, 
and $|S^v_{t_*}|$ is the number of actions taken by $v$ before time $t_*$.
$f(\cdot)$ is the joint probability density function of the time sequence 
 $T(S^v_{t_*})$ and $ |S^v_{t_*}|$.

As we show in Appendix \ref{MLE}, by taking the partial derivative of $log\,L(\alpha,\tau)$ with respect to $\alpha$ , we get the estimate values for $\alpha$ :
\begin{equation}
\hat{\alpha} =  \frac{|S_{t_*}|}{\sum_{v\in V} H_v(t_*,\tau)}
\label{eq:alpha}
\end{equation}

Fixing $\alpha$ to $\hat{\alpha}$, we get 
\begin{equation}
  \begin{aligned}
\hat{\tau} &= argmax [|S_{t_*}|log( \frac{|S_{t_*}|}{ \sum_{v\in V} H_v(t_*,\tau) }) - |S_{t_*}| \\
&+ \sum_{(v_i,t_i)\in S_{t_*}} log(h_{v_i}(t_i,\tau))]
\end{aligned}
\label{eq:MLE_tau}
\end{equation}
where $h_v(t,\tau)$ and $H_v(t,\tau)$ are introduced for simplicity's sake:

\begin{equation}
h_v(t,\tau) =  \sum_{(v_i,t_i)\in S_t}(prox(v_i,v)\,e^{-(t-t_i)/\tau})
\label{eq:f}
\end{equation}
and
\begin{equation}
H_v(t,\tau) = \tau \sum_{(v_i,t_i)\in S_t}(prox(v_i,v) (1 - e^{-(t-t_i)/\tau}))
\label{eq:F}
\end{equation}


Due to the complexity of $H(t,\tau)$ and $h(t,\tau)$ with regards to $\tau$,
it is not possible to obtain a closed-form solution for the maximum-likelihood estimate of $\tau$. 
However, since $\tau$ is the only variable here, we can use any line search algorithm to find the $\hat{\tau}$,
and techniques such as simulated annealing can be adopted to avoid falling into local optimal value.

\subsection{Prediction Phase}
After we learn the parameters $\tau$ and $\alpha$, we can generate the prediction of the action sequence after $t_*$.
To do this, we keep track of the next action of each user in a list in the time order.
Every time we pull the earliest action from the list and add it to prediction,
then we update the list to capture the future actions that will be triggered by this action.
The procedure is as follows:

\renewcommand{\theenumii}{\roman{enumii}}
\renewcommand{\labelenumii}{\theenumii}
\begin{enumerate}[noitemsep,nolistsep]
\item Calculate  $r_v(t_*)$, the activeness at time $t_*$ for each node $v$ in the network, using Equation \ref{eq:r}.

\item Generate a next action for each node in the network from the pdf in Equation \ref{eq:waiting_time}. 
			Sort the actions in the ascending order of time, store them in a list $L$. 

\item While $L$ is not empty, pull the first action $(v_i,t_i)$ from it.
		\begin{enumerate}[noitemsep,nolistsep]
				\item if $t_i>t_{end}$, jump to step 4.
				\item Add $(v_i,t_i)$ to $S'$, the predicted sequence.
				\item For all the nodes that are reachable from $v_i$, update their activeness $r_v(t_i)$ by Equation \ref{eq:r}.
				\item Update the next action time using the pdf in Equation \ref{eq:waiting_time}, and sort the list $L$ again.
		\end{enumerate}
\item Calculate \textbf{intensity}, \textbf{coverage}, \textbf{duration} using the predicted sequence $S'$.
\end{enumerate}

$t_{end}$ in Step 3-i is the end point of the last time interval on which we want to predict the trend. 
Notice that the sequence generating process is a random process. 
We may repeat Step 3 several times to get the average value of aggregates.
If implemented in a naive manner, the calculation of $r_v(\cdot)$ can be very time-consuming.
We will describe an efficient implementation in Section \ref{sec:effi}.


\subsection{Efficient Implementation}
\label{sec:effi}
As a property of the DA model, the effect of the sequence of actions can be considered as the sum of the individual effects by each single action. 
We call it \textbf{stacking principle}.
We can reduce the complexity of prediction algorithm based on the stacking principle. 
The general idea is to transform a complicated formula to a summation over the sequence of actions, which is easier to calculate.
The principle can be applied to both the parameter learning phase and prediction phase.
We are going to show the details of the implementation in Sections \ref{sec:learn_time} and \ref{sec:pred_time},
and discuss the time complexity in Section \ref{sec:time}.


\subsubsection{Speedup of Parameter Learning Phase}
\label{sec:learn_time}
For the parameter learning phase, the main complexity of computation comes from the calculation of $\sum_{v\in V} H_v(t_*,\tau)$ which is in both Equations \ref{eq:alpha} and \ref{eq:MLE_tau}. 
It will be very inefficient if we calculate the summation over all the nodes in the network in every step of the search algorithm. 
To reduce the complexity of calculation, we use the following equation:
\[
\begin{aligned}
\sum_{v\in V} H_v(t_*,\tau)=& \sum_{v\in V} [\tau \sum_{(v_i,t_i)\in S_{t_*}}(prox(v_i,v) (1 - e^{-(t_*-t_i)/\tau}))]\\
=& \tau \sum_{(v_i,t_i)\in S_{t_*}} [(1 - e^{-(t_*-t_i)/\tau}) \sum_{v\in V}prox(v_i,v)]\\
\end{aligned}
\label{eq:MLE_tau_2}
\]
The summation over all the nodes $\sum_{v\in V}prox(v_i,v)$ in the last equation does not depend on $\tau$.
It implies that for each $v_i$ that $(v_i,t_i) \in S_{t_*}$, we can calculate the summation $\sum_{v\in V}prox(v_i,v)$ once and store it.
Then in each step we only need to take a summation over all the actions in the trend sequence $S_{t_*}$.
Notice that the number of actions in  $S_{t_*}$ is usually much smaller compared to the total number of nodes in the graph.
Therefore, it is much more efficient to be calculated than the original form.

\subsubsection{Efficient Implementation of Prediction Phase}
\label{sec:pred_time}
For the prediction phase, the action rate function $r_v(t)$ in Equation \ref{eq:r} is a summation over all the actions in the network before time $t$.
According to the property of Poisson processes, we can generate an action sequence for each action $(v_i,t_i)$ before time $t$ with the action rate function $\alpha\cdot prox(v_i,v) e^{-(t-t_i)/\tau}$,
and then adding all these sequences together to get the action sequence for $v$.
In other words, instead of calculating $r_v(t)$ for node $v$ repeatedly, we can use the follow procedure to get exactly the same result: 
(1) For the action $(v_i,t_i)$, 
generate the sequence of actions that is caused by this single action for every  node reachable from $v_i$, 
(2) and then merge this sequence of actions with the current list of action to generate the new list. 
This stacking principle greatly simplifies the prediction algorithm.


\subsubsection{Time complexity}
\label{sec:time}
Due to the use of stacking principle, we can reduce the time complexity of the algorithm.
Let the values of $\alpha$ and $\tau$ be $n_p$-digit precision, and $m_{prox}$ be the maximum number of nodes in the proximity of a node, $|S|$ be the total number of actions in the sequence $S$.
In the learning phase, calculating the intermediate result takes $O(m_{prox}\cdot|S|)$.
The total number of steps of line search is $O(n_p)$, and each step of line search takes $O(|S|)$ to calculate the new value of objective function.
Thus, the time complexity of learning phase is $O(n_p\cdot|S|+m_{prox}\cdot|S|)$.
In the prediction phase, for each action in the $S$, the algorithm takes $O(|S|)$ time to generating the prediction sequence.
The time complexity of prediction phase is $O(m_{prox}\cdot|S|)$.
Thus, the total complexity is $O(n_p\cdot|S|+m_{prox}\cdot|S|)$. 
Usually, $n_p$ is much smaller than $m_{prox}$, we can think the time complexity as $O(m_{prox}\cdot|S|)$.

The time complexity of a naive algorithm that does not make use of the stacking principle is $O(n_p\cdot m\cdot|S|+ m \cdot |S|^2 )$. 
It is much slower than our algorithm.
The time complexity of prediction model based on the IC model is $O(m_{degree}\cdot|S|)$, where $m_{degree}$ is the maximum number of nodes that are directly linked to a node.
It is different from the time complexity of our algorithm in $m_{degree}$. 
That is because the IC model only considers the influence between nodes that are directly linked to each other.
Our algorithm has similar time complexity with the IC model, though the IC model is a much more simplistic model.

\section {Experiment}
\subsection{Algorithms and Performance Measures}
As we mentioned in Section \ref{sec:activeness}, for the DA model, we use two different measurements for the proximity between two nodes in the network.
For the shortest path measurement (DA-sp), we set the decay factor $b$ to $10$ in the experiment.
For the random walk measurement (DA-rw), the restart probability $p$ is set to $0.4$.

We compare the DA model with three variants of the widely used IC model. 
All of the three variants assume that each action comes with a delay, so that they can be used to model dynamic trends:
\begin{enumerate}
\item{\bf  eExp} (Edge-dependent exponential delay model \cite{Saito2009})  assumes that all the actions propagate through a certain edge are drawn independently from the same exponential distribution. 
\item{\bf tExp} (Trend-dependent exponential delay model  \cite{Saito2009})  assumes that delays for all the actions of a certain trend are drawn independently from the same exponential distribution.
\item{\bf tEqu} (Trend-dependent equal-length delay model  \cite{Leskovec2007})  assumes that there is a fixed-length delay for all the actions of a certain trend. 
\end{enumerate}
Parameters of the three baselines model are learned by the algorithm proposed in \cite{Saito2009}. 
For the intensity prediction, we extend the IC variants with a ``multiple actions factor'' to allow a node to perform actions more than once. 
These extended IC variants will be explained later in Section \ref{sec:intensity}.



To evaluate coverage and intensity, we use two measures: the error ratio and the coefficient of variation.

Error ratio is used to evaluate the goodness of the prediction compared with the true value. 
The formula of error ratio is given by:
\[
error\ ratio = \frac{|truth - prediction|}{truth}
\]

Since all of the tested algorithms are stochastic algorithms, we also evaluate the variance of the outputs. 
For every test, we run each algorithm for multiple times and estimate the coefficient of variation.
The coefficient of variation is estimated by:
\[
\hat{C_v} = \frac{s}{\bar x}
\]
where $s$ is the sample standard deviation, and $\bar{x}$ is the sample mean. 


\subsection {Trend Data for Evaluation}
As we described in Section \ref{sec:DataSets}, we use two social networks for the evaluation. 
For each network, we conduct experiments on multiple trends, as listed in Table \ref{tab:topics}.

For {\bf DBLP} dataset, we test 10 trends of hot keywords in the areas of data mining and machine learning.
For each trend, we use the trend sequence before year 2005 (2005 excluded) as the training sequence,
and the sequence from year 2005 to 2009 (2009 included) as the test sequence.
We take each year as a time interval, on which we consider intensity and coverage.

For {\bf Twitter} dataset, we test the algorithms on trends of 10 most popular hashtags in the last four months of year 2010.
For each of the trends, we use the trend sequence from the 40th week to 47th week as training sequence,
and the sequence from the 48th week to 52nd week as the test sequence.
We take each week as a time interval, on which we consider intensity and coverage.


\begin{table}[t]
\centering
	\scalebox{0.8}{
\begin{tabular}{|l l | l l|}
	\hline
	\multicolumn{2}{|c|}{\bf DBLP} & 	\multicolumn{2}{|c|}{\bf Twitter}\\
	\hline
	Concept drift&Boosting				&Sarcasm&Twibbon\\
	Kernel methods&Privacy					&SocialMedia& Wikileaks\\
	SQL&Face recognition 					&Geek&Awesome\\
	Heterogeneous network&Decision tree	 	&SoundCloud&Android\\
	Active learning&Streams 			&HTML5&Apple\\
	\hline
  \end{tabular}
	}
  \caption{Trends in the DBLP and Twitter dataset}	
  \label{tab:topics}
\end{table}

\subsection{Coverage}
\begin{figure}[t]
\centering
	\subfigure[\small DBLP]{\includegraphics[width=0.22\textwidth] {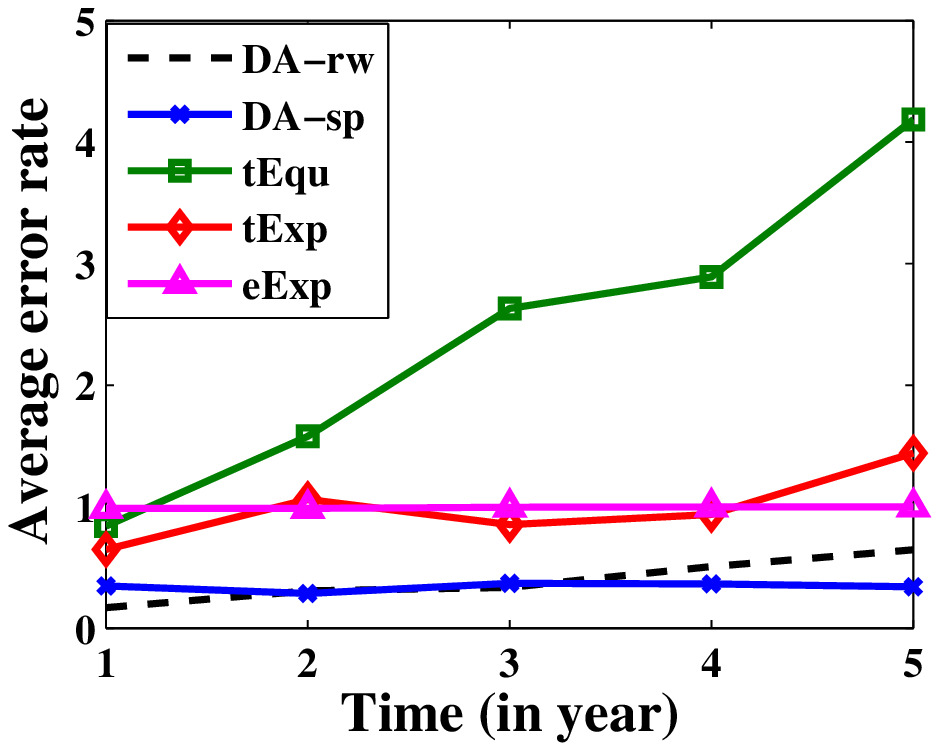} \label{fig:cover_error_dblp}}
	\subfigure[Twitter]{\includegraphics[width=0.22\textwidth] {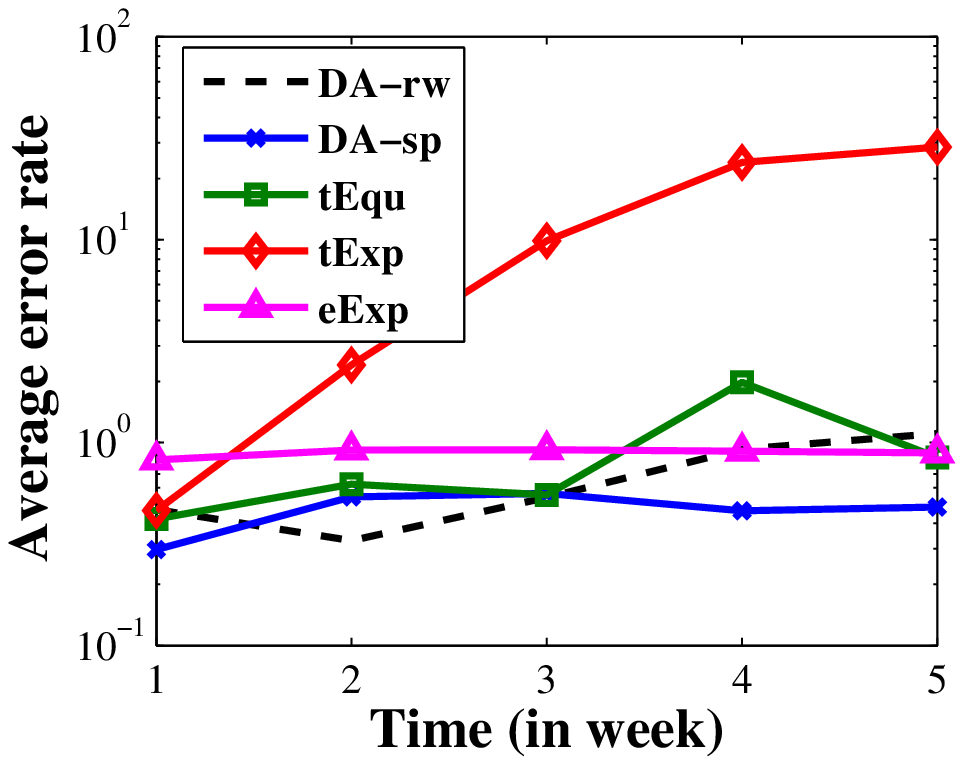}  \label{fig:cover_error_twitter}}
\caption{Error Ratio for Coverage Prediction}
\label{fig:cover_error}
\end{figure}

In Figure \ref{fig:cover_error}, we plot the average error ratio for 5 consecutive time intervals after the end point of training sequence.
As shown in the figures, error ratios for all the algorithms tend to increase as time progresses.
For the DBLP dataset, the error ratio of tEqu deteriorates very quickly, while for the Twitter dataset, tExp is the worst one.
In both figures, DA-rw and DA-sp have lower error ratios than the baselines for most of the ranges of X-axes, while eExp is in the middle.
The difference between the error ratio of DA-rw and the error ratio of DA-sp is not large,
which shows that the DA model is not sensitive to the different measurements of proximity, as long as the measurements are reasonable.

\begin{figure}[t]
\centering
\subfigure[\small DBLP]{\includegraphics[width=0.22\textwidth] {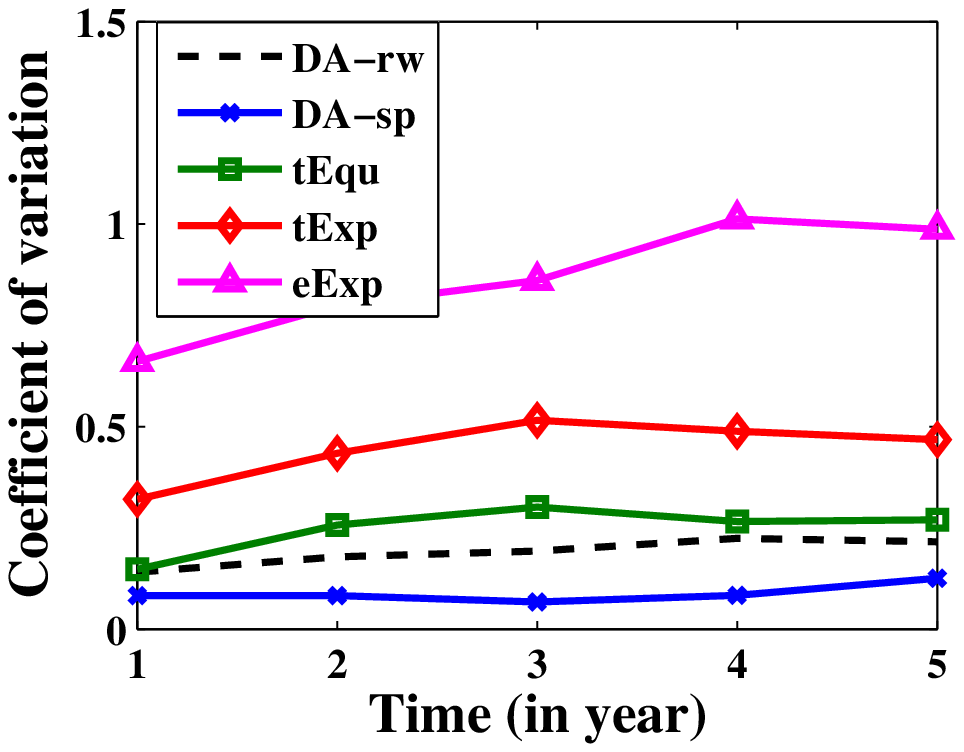} \label{fig:cover_cv_dblp}}
	\subfigure[\small Twitter]{\includegraphics[width=0.22\textwidth] {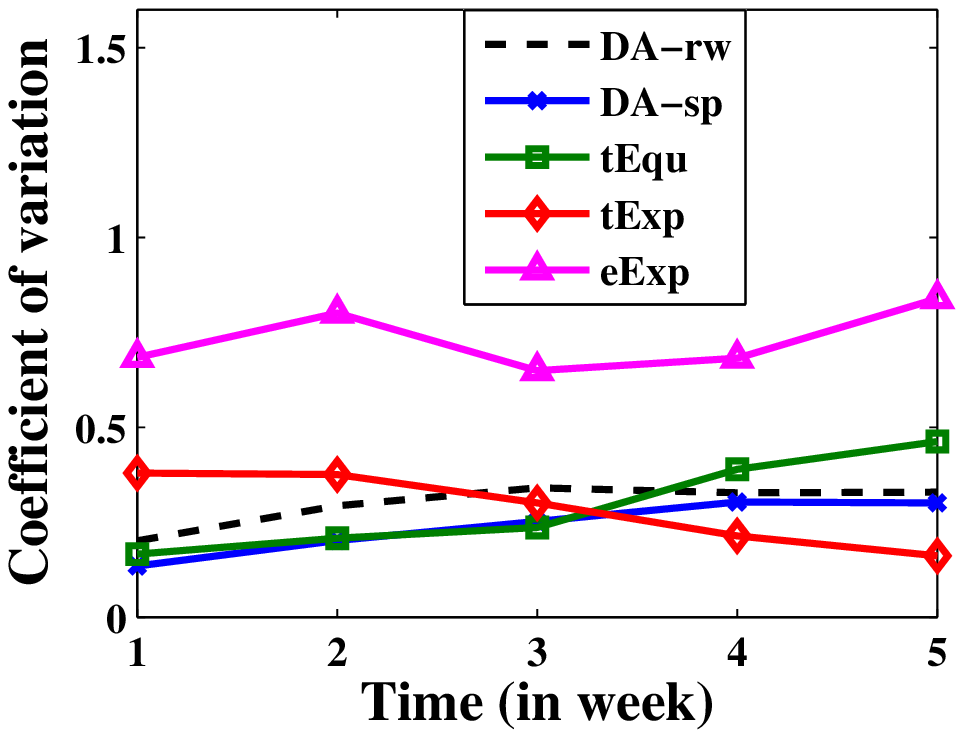}  \label{fig:cover_cv_twitter}}
\caption{Coefficient of Variation for Coverage Prediction}
\label{fig:cover_cv}
\end{figure}

In Figure \ref{fig:cover_cv}, we show the coefficient of variation for the coverage prediction.
For the DBLP dataset, DA-rw and DA-sp have a lower coefficient of variation than the baselines,
which means that the predictions got by the DA model are more stable.
For the Twitter dataset, the coefficient of variation of tEqu is similar to the DA model.
tExp has a strange decreasing curve, and its coefficient of variation is the lowest in the last two weeks of prediction.
Considering the large error ratio of tExp in later weeks in Figure \ref{fig:cover_error_twitter}, 
its low coefficient of variation is most likely to be caused by a large mean value of its prediction.
Thus, it is not comparable with the other three algorithms.
Among all the algorithms, eExp have highest coefficient of variation, hence it is less stable than all the other algorithms.
On both of the datasets, DA-sp has a slightly lower coefficient of variation than DA-rw, which mean that the shortest path distance measurement makes the result more stable than the random walk measurement.

\subsection{Intensity}
\label{sec:intensity}

\begin{figure}[t]
\centering
	\subfigure[\small DBLP]{\includegraphics[width=0.22\textwidth] {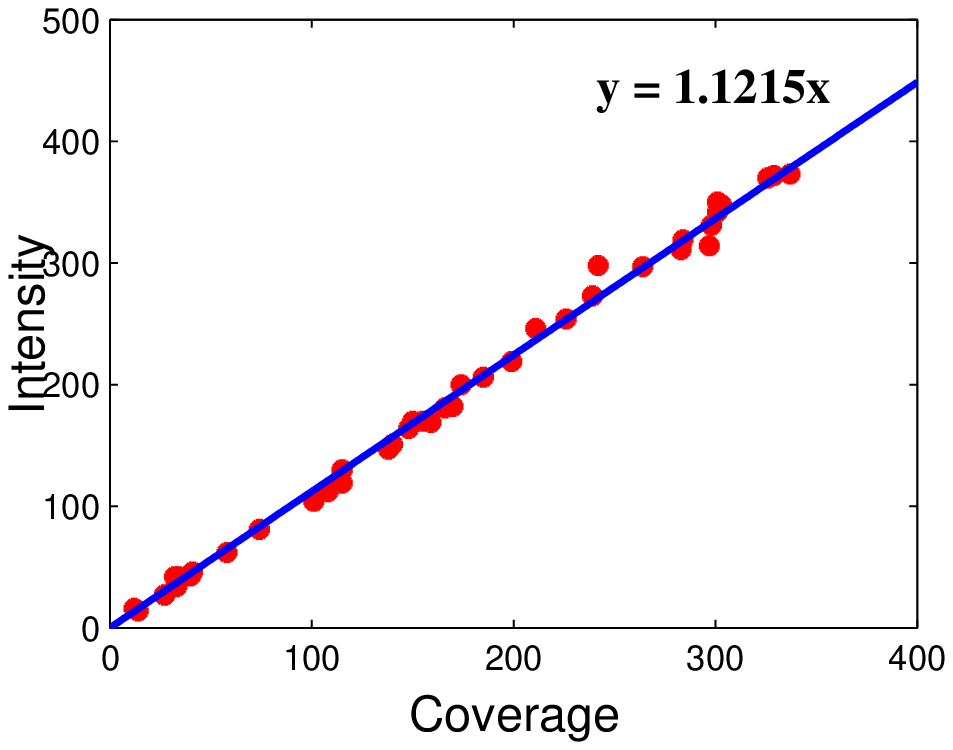} \label{fig:cover_intensity_dblp}}
	\subfigure[\small Twitter]{\includegraphics[width=0.22\textwidth] {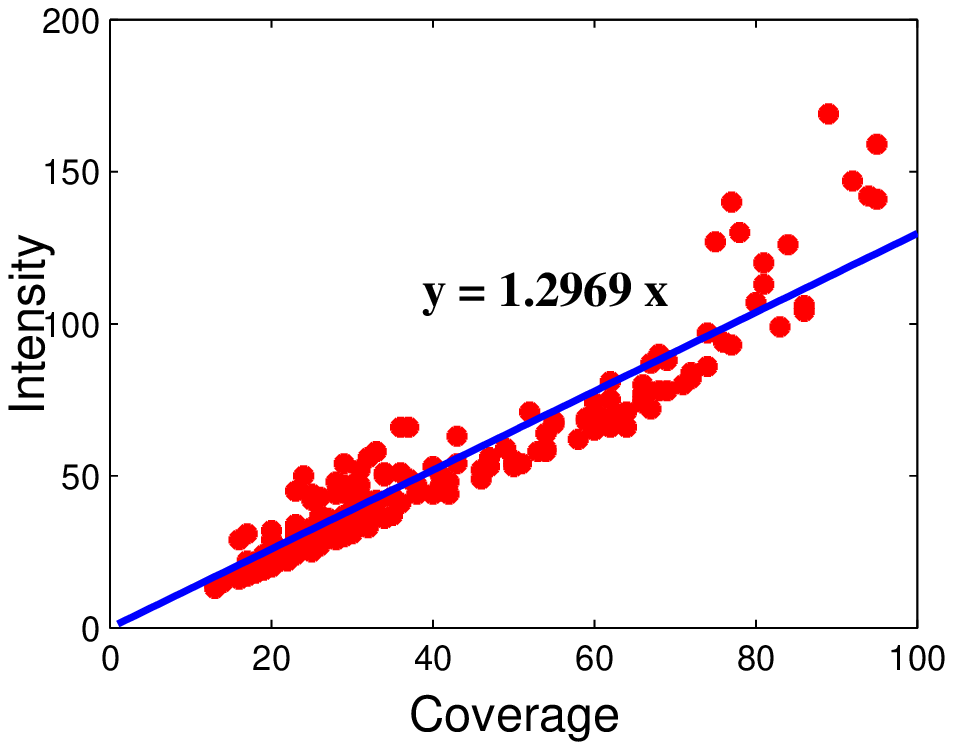}  \label{fig:cover_intensity_twitter}}
\caption{Relationship between Intensity and Coverage}
\label{fig:cover_intensity}
\end{figure}

For the evaluation of intensity prediction,
we use tEqu-mult, tExp-mult and eExp-mult instead of tEqu, tExp, and eExp as baselines.
It is because the tEqu, tExp and eExp, like the standard IC model,
do not allow multiple actions being performed by the same node.
To construct better baselines, we try to capture the relationship between the coverage and intensity. 
Figure \ref{fig:cover_intensity} shows our observation on the relationship.

For each dataset, we calculate coverage and intensity for all the intervals in the training time, and plot each pair of coverage and intensity in this Figure \ref{fig:cover_intensity}.
In both figures, the values of coverage are illustrated on the X-axes, and the values of intensity are illustrated on the Y-axes.
As shown in Figure \ref{fig:cover_intensity}, for the DBLP trends, the proportional function $y = 1.1215x$ fits the relationship of intensity and coverage quite well,
while for the Twitter trends, the $y = 1.2969x$ fits the relationship.

\begin{figure}[t]
\centering
	\subfigure[\small DBLP]{\includegraphics[width=0.4\textwidth] {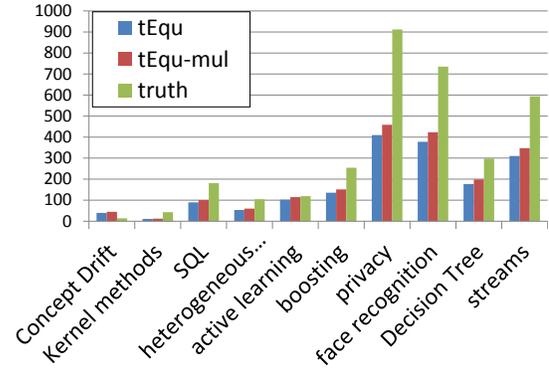} \label{fig:mult_dblp}}
	\subfigure[\small Twitter]{\includegraphics[width=0.4\textwidth] {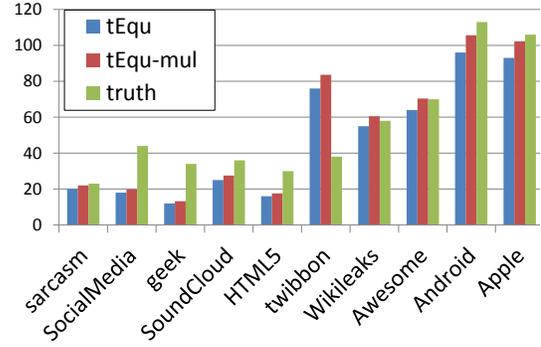}  \label{fig:mult_twitter}}
\caption{Comparation of tEqu and tEqu-mult}
\label{fig:mul}
\end{figure}

Based on this observation, we add a multiple action factor to the three baselines and get three new baselines.
Figures \ref{fig:mult_dblp} and \ref{fig:mult_twitter} show the improvement of tEqu-mult from tEqu on the DBLP and Twitter datasets respectively.
For each trend, we show the predictions of intensity made by tEqu-mult and tEqu for in the first prediction interval (year 2005 for the DBLP trends and the 48nd week for the Twitter trends) as well as the true value.
As illustrated in the figure, tEqu tends to make a prediction lower than the true value because it does not allow multiple actions taken by each node.
tEqu-mult makes a better prediction by adding a multiple factor.

\begin{figure}[t]
\centering
	\subfigure[\small DBLP]{\includegraphics[width=0.22\textwidth] {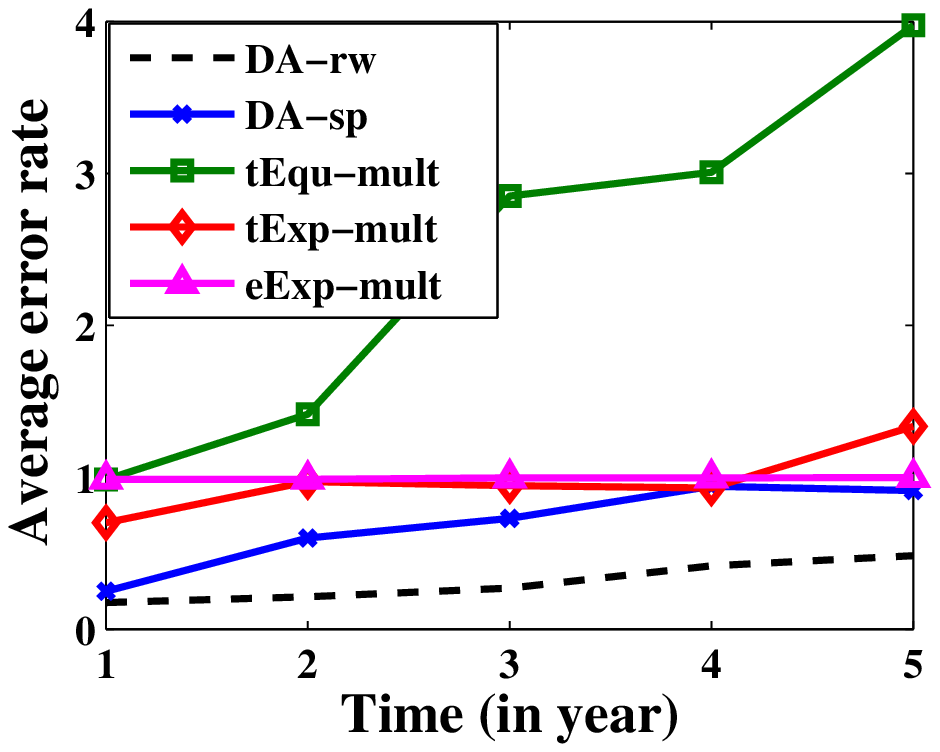} \label{fig:int_error_dblp}}
	\subfigure[\small Twitter]{\includegraphics[width=0.22\textwidth] {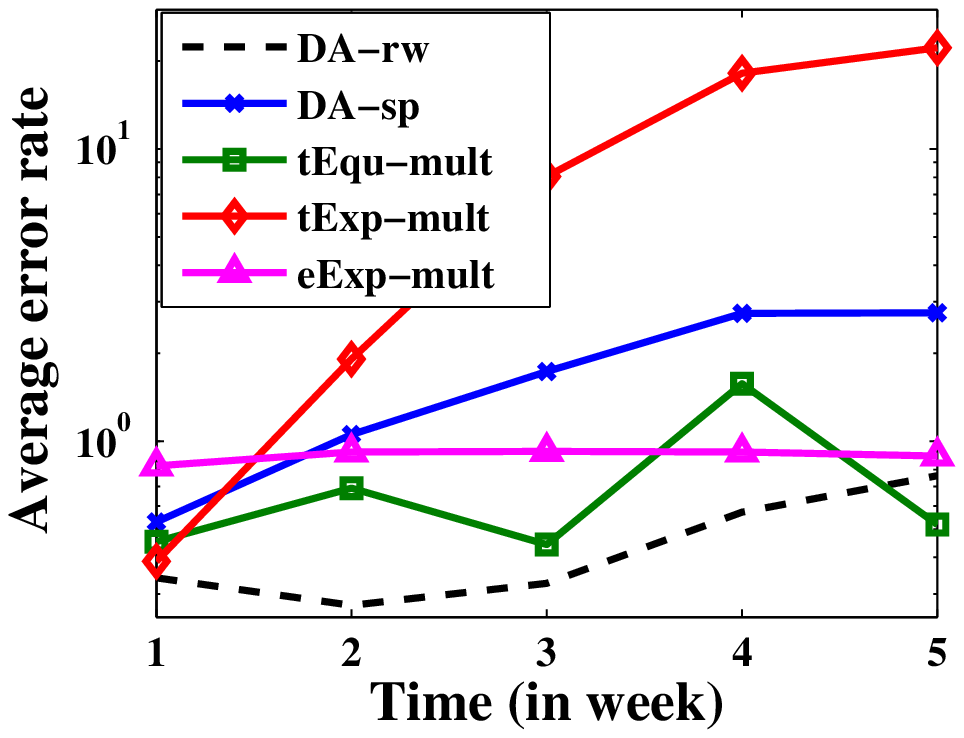}  \label{fig:int_error_twitter}}
\caption{Error Ratio for Intensity Prediction}
\label{fig:int_error}
\end{figure}

Figure \ref{fig:int_error} shows the error ratios for intensity prediction.
As shown in the figures, for all the algorithms, error ratios tend to increase as time progresses.
Among the four algorithms, tEqu-mult has the highest error ratio for the DBLP dataset, while tExp-mult is the worst one for the Twitter dataset.
DA-sp performs well on the DBLP dataset, but has a high error rate on the Twitter dataset. 
That may due to the relatively smaller graph diameter of the Twitter dataset. 
DA-rw has lower error ratios than other algorithms for most of the ranges of the X-axes for both datasets.

\begin{figure}[t]
\centering
	\subfigure[\small DBLP]{\includegraphics[width=0.22\textwidth] {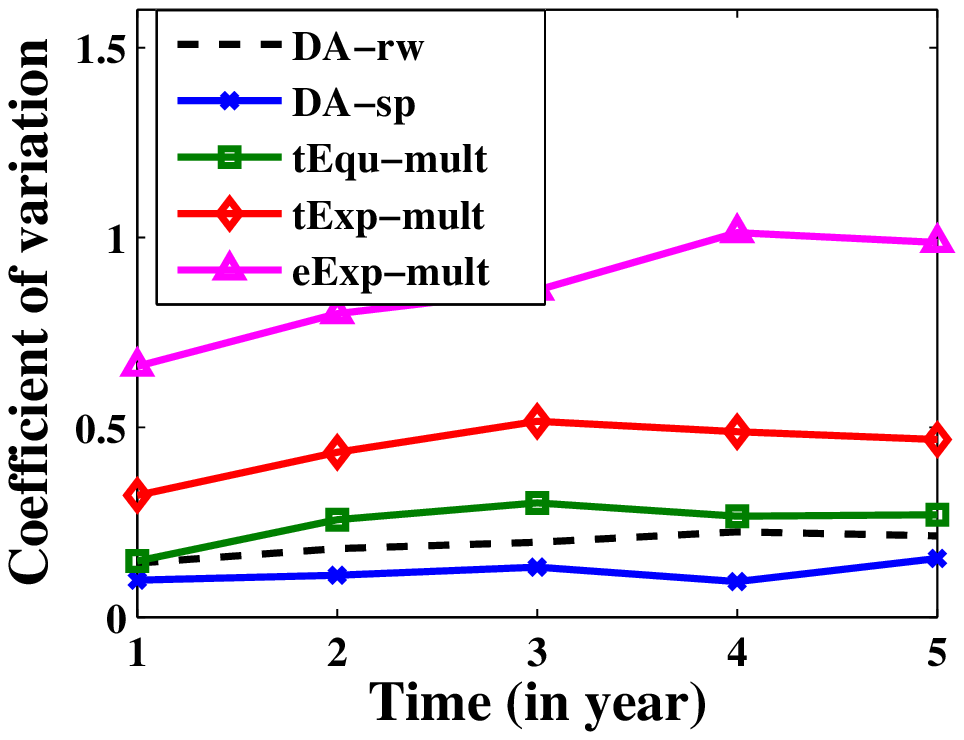} \label{fig:int_cv_dblp}}
	\subfigure[\small Twitter]{\includegraphics[width=0.22\textwidth] {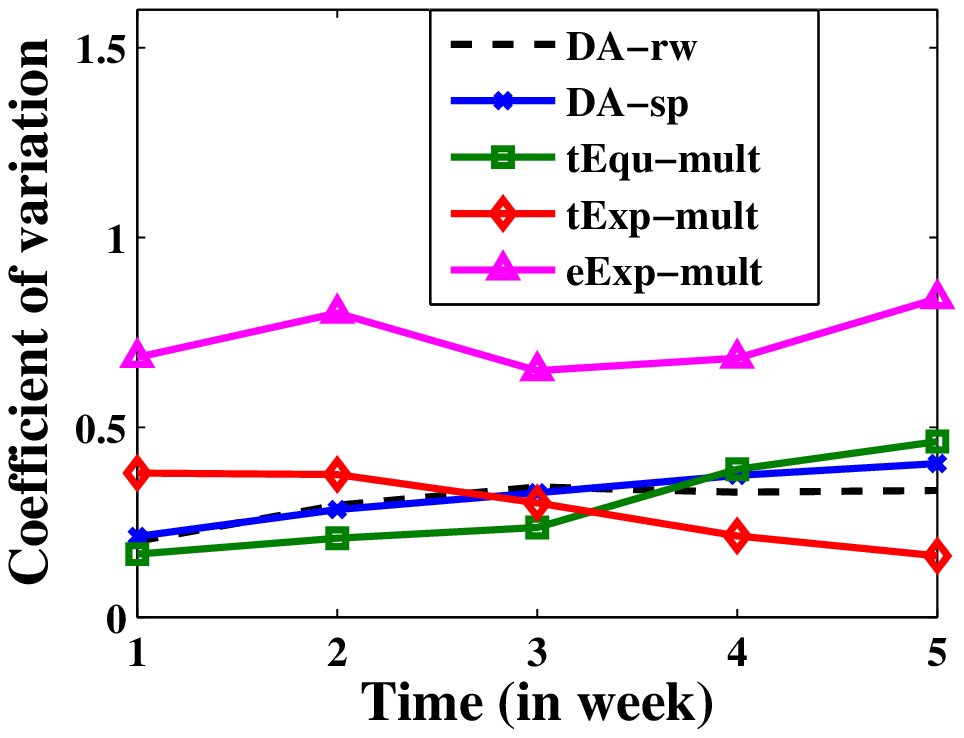}  \label{fig:int_cv_twitter}}
\caption{Coefficient of Variation for Intensity Prediction}
\label{fig:int_cv}
\end{figure}

In Figure \ref{fig:int_cv}, we show the coefficient of variation for the intensity prediction.
Notice that the result is similar to the coefficient of variation for coverage prediction in Figure \ref{fig:cover_cv}.
It is because the result of tExp-mult is proportional to the result of tExp, their coefficients of variation should be the same. 
Therefore, the curves for tExp-mult in Figure \ref{fig:int_cv} are exactly the same as the curves for tExp in Figure \ref{fig:cover_cv}.
So are the curves for tEqu-mult in Figure \ref{fig:int_cv} and curves for tEqu in Figure \ref{fig:cover_cv}.
For the DBLP dataset, our algorithms have lower coefficient of variation than the baselines.
For the Twitter dataset, the coefficient of variation of tExp-mult drops as time progresses, as a result of its large and inaccurate prediction for intensity,
while the curves for DA-rw, DA-sp and tEqu-mult are similar.
For both of the datasets, the prediction of eExp is less stable than the other four algorithms.

\subsection{Duration}
We test both coverage-based duration and intensity-based duration.
To calculate the duration, we use the coverage and intensity at the last observed interval (year 2004 for the DBLP dataset, the 47th week for the Twitter dataset) as thresholds.
Since the length of the test time is limited, coverage or intensity for about half of the trends never drops below the thresholds.
We make each algorithm predict whether the duration covers all the 5 prediction intervals, 
and report the accuracy of this prediction.

Table \ref{tab:duration} shows the accuracy of the duration prediction.
In the table, ``C-D'' and ``I-D'' are short for ``Coverage-based Duration'' and ``Intensity-based Duration''.
As shown in the table, DA-rw makes the best predictions of accuracy in three of the four cases. 
The accuracy of DA-sp and tExp is similar expect for the last case.
tEqu has the lowest accuracy on the DBLP dataset, while eExp has lowest accuracy on Twitter dataset.

\begin{table}[ht]
\centering
		\scalebox{1.0}{
	\begin{tabular}{ l  l  l  l  l }
	\hline
	&\multicolumn{2}{c}{DBLP}&\multicolumn{2}{c}{Twitter}\\
	\hline
	&C-D&I-D&C-D&I-D\\
	\hline
	DA-rw& 0.9 & 0.9 & 0.8 & 0.5\\
	DA-sp& 0.8 & 0.9 & 0.6 & 0.5\\
	tEqu(-mult)& 0.6 &  0.5 & 0.4 & 0.5\\
	tExp(-mult)& 0.8 &  0.9 & 0.6 & 0.7 \\
	eExp(-mult)& 0.7 & 0.7 & 0.3 & 0.4\\
	\hline
  \end{tabular}}
  \caption{Accuracy of Duration Prediction}	
  \label{tab:duration}
\end{table}

\subsection{Case Study}
We select two trends for the case study: the trend for ``android'' in the Twitter network and the trend for ``concept drift'' in the DBLP network.

\begin{figure}[t]
	\centering
	\subfigure[\small Android]{\includegraphics[width=0.22\textwidth] {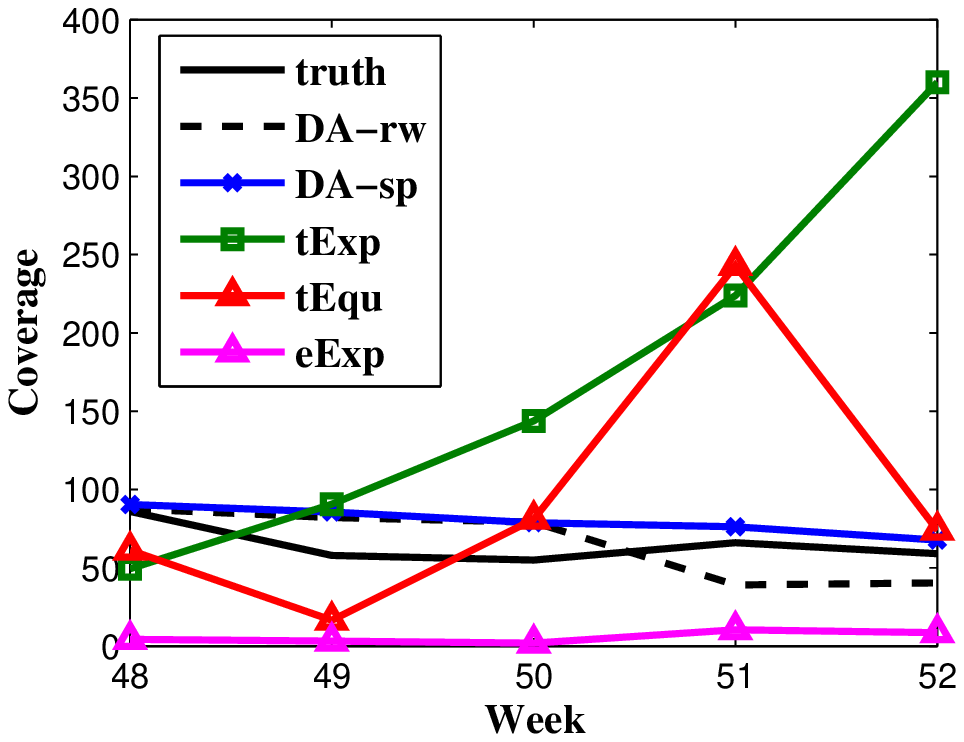} \label{fig:android}}
	\subfigure[\small Concept drift]{\includegraphics[width=0.22\textwidth] {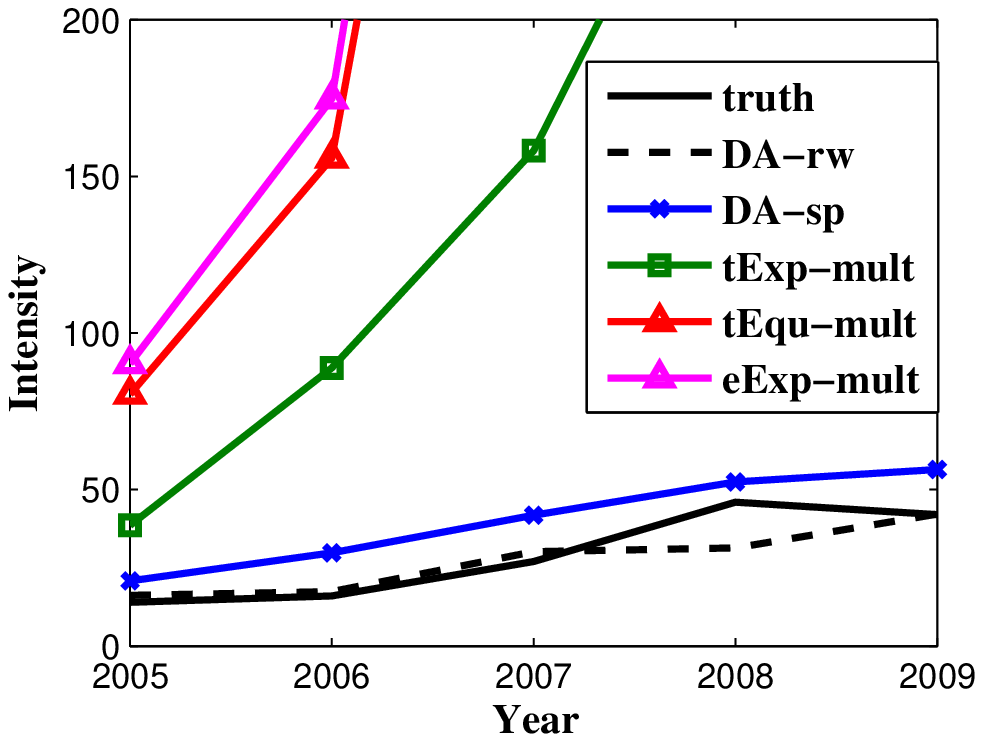} \label{fig:concept_drift}}
\caption{Case study on two trends in DBLP and Twitter dataset}
\label{fig:case_study}
\end{figure}

In Figure \ref{fig:android}, we show the predicted and true coverage of the trend ``android'' in the last five weeks of 2011.
The coverages predicted by DA-rw and DA-sp are close to the true value.
Besides, only these two curves descend gently as the curve of true coverage.
The coverage predicted by tExp quickly becomes much larger than the true value, while the coverage predicted by eExp is always very small.
The coverage predicted by eExp fluctuates dramatically.
If we choose $\theta = 86$ (the true coverage value in week 47),
the coverage-based duration predicted by DA-rw, DA-sp will be 1 week, the same as the true value,
while the duration predicted by three baselines will all be 0 week, since the predicted value falls below $\theta$ in the week 48. 

In Figure \ref{fig:concept_drift}, we show the predicted and true intensity of the trend ``concept drift'' in the years 2005-2009.
All the curves keep increasing during the five intervals.
But the intensity predicted by DA-rw and DA-sp is much closer to the true value than the value predicted by the baselines.

\subsection{Variations of Parameters}
\label{sec:variation_of_para}

\begin{figure}[htb]
\centering
	\subfigure[]{\includegraphics[width=0.2\textwidth] {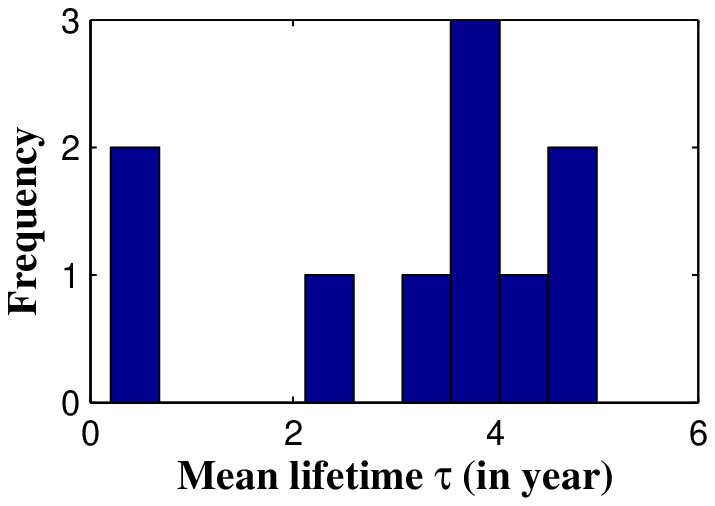} \label{fig:dist_alpha}}
	\subfigure[]{\includegraphics[width=0.2\textwidth] {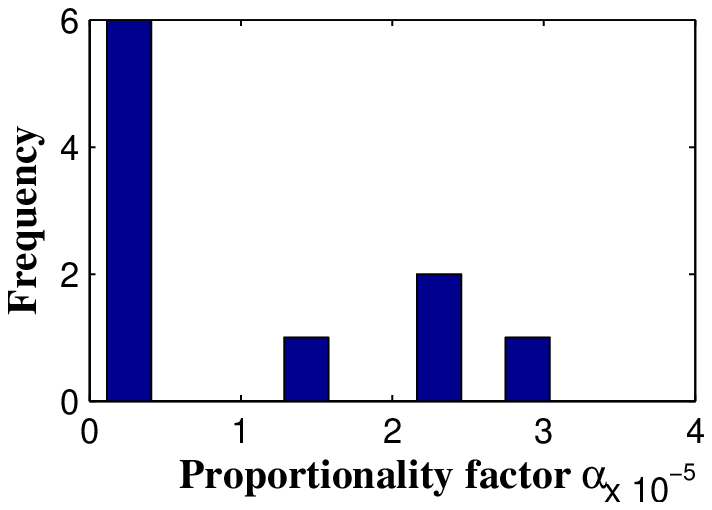}  \label{fig:dist_tau}}
\caption{Distribution of Parameters}
\label{fig:dist}
\end{figure}

Figure \ref{fig:dist} shows the distributions of the parameters, mean life time $\tau$ and proportionality factor $\alpha$, for DBLP data set.
As shown in the figure, each of the parameters has a large variation over trends.
For example, most of the trends have $\tau$ (mean lifetime of activeness decay) around 4 years, 
but some trends have very small $\tau$ which is less than 1 year.
The large variations show that the cascade processes for trends are different from each other in nature.
As a result, it is necessary to learn the model for each individual trend.

\subsection{Summary and Discussion}
We summarize the result as follows:
For the prediction of coverage, both DA-rw and DA-sp have better performance than all the three baselines.
For the prediction of intensity and duration, DA-rw works better than the baselines on both datasets.
while DA-sp makes better predictions than baselines on the DBLP dataset.
For all the evaluations, DA-rw has comparable or better results than DA-sp.

The two datasets we used are very different in many aspects. 
For example, the time granularity of the DBLP dataset is one year,
while in the Twitter dataset the time granularity is one millisecond.
The DA model can make accurate predictions on both of dataset.
It shows that the DA model is practical for various applications.

According different applications, it may not always be necessary to make predictions on all of the three measures.
As shown in Figure \ref{fig:cover_intensity}, 
the linear relationship between intensity and coverage is very significant in the DBLP dataset,
comparing to that of the Twitter dataset.
That is because the intensity of trends in the Twitter dataset can easily be spammed by a small fraction of users,
but it is much harder to publish a paper than posting a tweet.
Therefore, for applications on the DBLP network, 
the measures of intensity and coverage are roughly interchangeable. 
For the DBLP dataset, we may only need to predict one of them.
Nevertheless, each of the three measures is useful for some applications, 
and it is desirable to propose a model that can capture all of them at the same time. 


\section {Related Work}
Information diffusion processes in social networks have been intensively studied in \cite{Kempe2003,Goyal2010,Leskovec,Leskovec2007,Chen2010,Gruhl2004,Kossinets2008}. 
Most of this work considers trends in social network as the results of information diffusion processes.
These papers focus on predicting user-level behaviors rather than the aggregated measures of trends.
They usually model trends on a discretized time, which makes them not practical for predicting dynamic properties of trends.
Independent Cascade (IC) model and its variants are the most widely studied models for information diffusion processes \cite{Kempe2003,Leskovec2007,Chen2010}.
Most of this work assumes that the models are given as an input and try to solve the influence maximization problem on the model.
The work in \cite{Saito2008,Saito2009,Goyal2010, Dickens2012} focuses on learning the diffusion probabilities of the cascade models.
The work in \cite{Saito2009} defines a variant of the IC model that models the diffusion delay as a random variable with an exponential distribution, and provides an inference algorithm for the model.
The work in \cite{Wang2012} proposes a microscopic social influence model,
in which influence is modeled as ``heat'' that flows through the network.
It is conceptually resemble to the activeness in our model.
But the approaches are fundamentally different, 
and their model focuses on the prediction on each individual node.

Some existing work argues that the information diffusion inside social networks is not the only explanation of trends.
Homophily is also considered to be important for the modeling of trends \cite{Shalizi2011,Budak2011,Anagnostopoulos2008},
and the external influence from outside the network is studied in \cite{Myers2012}.

Recently, some work studies real-life trends in online communities.
Most of this work focuses on analyzing observed trends rather than predicting future trends,
and does not make use of the structure of the social network.
The work in \cite{Romero2011} studies the dynamic of trends on the Twitter community.
The work in \cite{Leskovec} analyzes the cascade behaviors on blogspace,
and the work in \cite{Leskovec2009} analyzes trends on news website and blogs.
The work in \cite{Budak2011} provides a technique to identify popular trends,
which utilizes the structural information of social networks.

\section {Conclusions}
This paper studies trends in social networks.
We identify coverage, intensity and duration as the three characteristics of a trend.
Though the phenomenon of trends has been widely observed and studied,
none of the previous models can capture all the three important aspects of trends.
We proposed a Dynamic Activeness model for trends based on the novel concept of node activeness.
The experimental result shows that the proposed DA model can predict trends more accurately than information diffusion models.

\bibliographystyle{abbrv}
\bibliography{trend_SDM}

\appendix
\section{Maximum Likelihood Estimation}
\label{MLE}

The joint probability density function is given by \cite{Zhao1996}:
\begin{equation}
 \textstyle{f(T(S^v_{t_*}),|S^v_{t_*}|;\alpha,\tau) =  e^{-\int_0^{t_*}r_v(t)\,dt} \cdot \prod_{t_i\in T(S^v_{t_*})}r_v(t_i)}
\label{eq:jdf}
\end{equation}


For simplicity's sake, we introduce $H_v(\cdot)$ and $h_v(\cdot)$ as Equations \ref{eq:f} and \ref{eq:F}.
Take Equations \ref{eq:jdf}, \ref{eq:f} and \ref{eq:F} into Equation \ref{eq:L1}, we get:

\[
  \begin{aligned}
L(\alpha,\tau) &= \textstyle{\prod_{v\in V} (e^{-\alpha H_v(t_*,\tau)} \cdot \prod_{t_i \in T(S^v_{t_*})}\alpha h_v(t_i,\tau))}\\
&= \textstyle{\alpha^{|S_{t_*}|} \prod_{v\in V} e^{-\alpha H_v(t_*,\tau)} \cdot \prod_{(v_i,t_i)\in S_{t_*}}h_{v_i}(t_i,\tau)}
\end{aligned}
\]

The log-likelihood function is then given by:
\[
  \begin{aligned}
	log\,L(\alpha,\tau) &= \textstyle{|S_{t_*}|log\alpha - \alpha \sum_{v\in V} H_v(t_*,\tau)}\\ 
			    &+ \textstyle{\sum_{(v_i,t_i)\in S_{t_*}} log(h_{v_i}(t_i,\tau))}
\end{aligned}
\]
By taking the partial derivative of $log\,L(\alpha,\tau)$, we get the maximum-likelihood estimation of $\alpha$ and $\tau$.
Notice that in the above inference we assume that the $\epsilon$ in Equation \ref{eq:r} is small enough and negligible.

\end{document}